\documentclass[a4paper,11pt]{article}
\pdfoutput=1 

\usepackage{jinstpub} 

\usepackage{lineno}
\usepackage{multirow}
\usepackage{upgreek}

\title{\boldmath Study on U/Th residual radioactivity in acrylic from surface treatment}

\author[a]{Yuanxia Li,}
\author[a]{Xiaohui Qian,}
\author[a]{Xiaolan Luo,}
\author[a,1]{Jie Zhao,\note{Corresponding author.}}
\author[b]{Gaofeng Zhang,}
\author[a]{Xiaoyan Ma,}
\author[a]{Yuekun Heng,}
\author[a]{Liangjian Wen,}
\author[c]{Monica Sisti,}
\author[d]{Fr\'{e}d\'{e}ric Perrot,}
\author[b]{Hongqiang Tang}

\affiliation[a]{Institute of High Energy Physics, Chinese Academy of Sciences, Beijing 100049, China}
\affiliation[b]{Donchamp New Material Technology Co. Ltd, Taixing 225400, China}
\affiliation[c]{INFN Milano Bicocca and Universit\`{a} di Milano-Bicocca, Milano, Italy}
\affiliation[d]{Univ. Bordeaux, CNRS, CENBG, UMR 5797, F-33170 Gradignan, France}

\emailAdd{zhaojie@ihep.ac.cn}

\abstract{Acrylic is widely used as material for the target container in low background experiments due to its high light transparency and low intrinsic radioactivity. However, its surface can be easily contaminated during production, so careful treatment of the surface is essential to avoid direct contamination of the target. The Jiangmen Underground Neutrino Observatory will use about 600~t of acrylic to build the spherical vessel of 35.4~m in diameter for a 20~kt liquid scintillator (LS). Since acrylic will contact the LS directly, the cleanliness of the its surface is quite important for the radiopurity of the LS. A new method for measuring the radioactivity of $^{238}$U and $^{232}$Th in acrylic to sub-ppt ($<10^{-12}$~g/g) was developed, and it is crucial for the acrylic radioactivity screening in this study. We performed many background tests on different surface treatments, and the recommended procedure for the treatment of acrylic to achieve low radioactivity and high light transparency could be applicable to other low background experiments.}

\begin{document}
\maketitle
\flushbottom

\section{Introduction}
The Jiangmen Underground Neutrino Observatory (JUNO) is a multi-purpose neutrino experiment, which will build the world's largest liquid scintillator (LS) detector. An acrylic vessel is used as the 20~kt LS container with 35.4~m in diameter and 120~mm in thickness because of its high light transparency and low intrinsic radioactivity~\cite{JUNO:2022hxd}. 

To reach the physics target, the average radio-purity of the acrylic is required to be less than 1~ppt~\cite{JUNO:2021kxb}. As shown in Ref.~\cite{Cao:2020zyr}, the radio purity in the bulk has reached the requirement. However, the distribution of radioactivity along the thickness is not extremely uniform. The bulk measurement of samples with several centimeters of thickness  can not reflect the surface contamination, since the thickness of external contamination on the acrylic surface usually reaches nanometre to micron level, and the surface contamination will result averaged out in the bulk measurement. If the external radioactive contamination stays on the surface, only  gammas and few alphas/betas from radioactive decays can be emitted into the LS and rejected by the effective fiducial volume cut by the off-line analysis. However, based on the experiences at Borexino~\cite{Borexino:2022pvu} and SNO+~\cite{SNO+leaching}, the U/Th daughters near the inner surface of the vessel could leak into the LS by diffusion and convection during stable running, thus making the acrylic vessel  a stable contamination source for the LS. Borexino put great efforts on controlling the temperature distribution to avoid convection in LS; however, it is impossible to repeat the same thing at JUNO due to the large size of the experiment (70 times Borexino target). Therefore, the radiopurity of the inner surface of acrylic is quite important for JUNO, especially for the physic topics in the low energy regions, such as solar neutrino studies~\cite{JUNO:2022hxd}. 

\begin{figure}[htbp]
\centering
    \includegraphics[height=5cm]{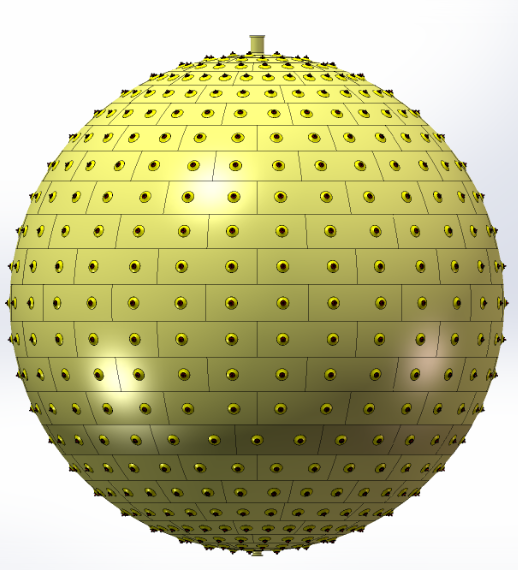}
    \includegraphics[height=5cm]{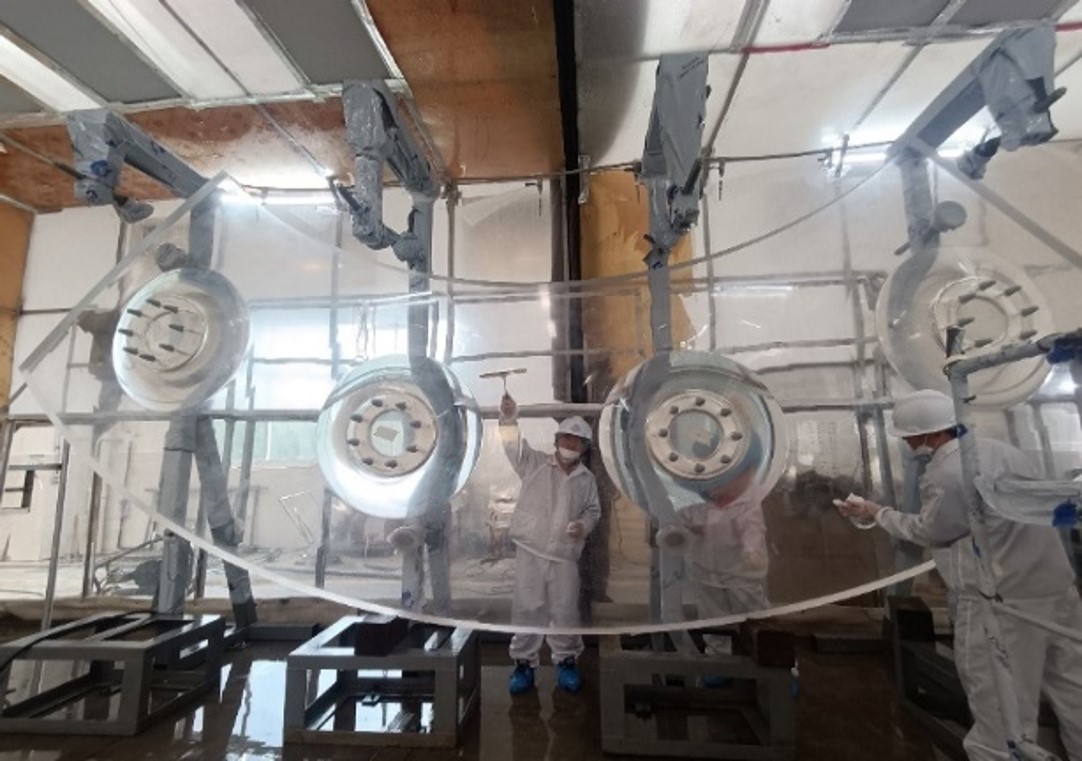}
    \caption{The acrylic vessel is produced with 265 pieces due to its very large size, as shown in the left figure. Each panel is produced separately, and all the panels will be finally bonded to a sphere at JUNO site. One JUNO panel is shown in the right figure as an example.}
    \label{fig:Acrylic_vessel}
\end{figure}

The JUNO acrylic vessel is produced with 265 pieces because of its large size, as shown in Fig.~\ref{fig:Acrylic_vessel}. Each acrylic panel is firstly produced in a carefully cleaned flat mold and then shaped to bend board at high temperature at the company; finally it is bonded directly at JUNO site~\cite{JUNO:2022hxd}. The surface of the acrylic panel is exposed to air and covered with a high-temperature resistant cloth during the bending, thus the surface is eventually contaminated and not as smooth as the flat panel. Surface treatments are performed before shipment to the JUNO site. Typical protocols to remove surface contaminations include the following steps: 

\begin{itemize}
\item The highest contaminations can reach tens of $\upmu$m depths from the surface~\cite{LA-ICPMS}. Thus, at least 0.1~$\upmu$m thickness of the surface of acrylic is firstly removed by sanding with special paper.

\item The surface is not so smooth after sanding, so polishing is needed to improve the transparency. The polishing is done with a wool wheel and polishing fluid.

\item The surface should be well cleaned with detergent and deionized water after polishing. 

\item Finally, a thin film is used to cover the acrylic surface to avoid external contamination during transportation and installation. 
\end{itemize}

All tools and detergents have risks to introduce more contamination on the surface, therefore each surface treatment should be well studied to minimize those risks. 

The paper is organized as follows. The surface treatment material screening is described in Sec~\ref{sec2}. The new method for measuring $^{238}$U and $^{232}$Th (U/Th) in acrylic to sub-ppt level is described in Sec~\ref{sec3}. Various tests on radioactivity from each step of the above treatment protocol are discussed in Sec~\ref{sec4}, and the recommended procedure is also shown in the same section. The concluding remarks are in Sec.~\ref{sec5}.

\section{Screening of surface treatment material by gamma spectrometer}
\label{sec2}
The main tools and detergents used for surface treatments are sanding papers, polishing fluid, Alconox powder~\cite{alconox}, deionized water and protection film. We used a gamma spectrometer to screen the radiopurity of the tools and detergents first. We have one HPGe facility at ground level at IHEP~\cite{Niu:2014xla}. The gamma spectrometer plays an important role in the primary screening of the raw materials at JUNO, and the advantage is that samples can be screened directly without any pre-treatment and with no damage to the sample. The principle of gamma spectroscopy is to measure the emitted gamma from the decay chain, and it is more sensitive to gammas with higher energy due to lower background. For example, the detector is more sensitive to the lower part of the uranium chain starting from $^{226}$Ra, and the $^{238}$U concentration can be derived by assuming secular equilibrium.

The measured results for the tools and detergent used for acrylic surface treatment are summarized in Table~\ref{tab:HPGe}. We found that special mirror wax purchased from Saint-Gobain~\cite{Norton} has lower radioactivity for $^{40}$K than one-step fast wax, due to the different chemical components. The sandpaper with a green base plate has lower radioactivity and better sand fastening than the one with a black base plate for similar mesh numbers. Regarding the Alconox powder, the first measurement performed at IHEP gave an upper limit of several tens to hundreds of ppb level for thorium and uranium concentrations, respectively, due to the limited sensitivity of HPGe. The same sample was further measured by HPGe at China Jinping Underground Laboratory (CJPL) with better shielding and sensitivity~\cite{Wang_2016} down to 10 ppb for $^{232}$Th. From the obtained results, there is a higher uranium concentration than thorium in the Alconox powder. The polishing fluid and sandpaper with lower radioactivity are selected for the surface treatments on acrylic. 

After surface treatment, the acrylic surface will be protected with a thin film to avoid dust and radon daughters deposition on the surface during the handling operations for transportation and installation. The film on the inner surface will be removed by the final cleaning with high pressure water jet after installation, and there is no possibility for manual operation inside the acrylic vessel. There are two types of protection film that can be used to protect the acrylic surface, ordinary polyethylene (PE) film with few glue on the surface and adhesive masking paper film. The PE film is better in toughness, but it is not easy to be removed by the water jet. On the contrary, the paper film with few water soluble glue on the surface can be easily removed by water washing since the glue is soluble in water, but overall its mechanical strength is worse. The radioactivity screening of the glue that is used on the PE film showed only upper limits for U/Th at the level of hundreds of ppb. Concerning the adhesive masking paper film (purchased from~\cite{film}), we could only measure the paper film and the glue together inside the HPGe detector. The water soluble glue itself was further measured by ICP-MS, as it will be discussed later.

\begin{table}[htbp]\small
\centering
 \caption{ \label{tab:HPGe} Material screening by gamma spectroscopy on tools and detergents used for the acrylic surface treatment. If not differently mentioned, the sample was measured at IHEP. The conversions between Bq/kg and mass concentration units for U/Th are given:  1~Bq/kg of $^{232}$Th activity in a material is equal to $2.5 \times 10^{-7}$\,g/g (or 250\,ppb) of $^{232}$Th mass concentration in that material; similarly, 1~Bq/kg of $^{238}$U means $8.1 \times 10^{-8}$\,g/g (or 81\,ppb) of $^{238}$U.}
 \smallskip
	\begin{tabular}{c|c|c|c|c|c}
        \hline
    \multicolumn{2}{c|}{} & \multicolumn{2}{c|}{$^{238}$U chain [Bq/kg]} & $^{232}$Th chain [Bq/kg] & $^{40}$K \\ 
    \cline{3-5}
    \multicolumn{2}{c|}{}& $^{214}$Bi/$^{214}$Pb & $^{226}$Ra & $^{212}$Bi/$^{208}$Tl & [Bq/kg] \\ \hline
    \multirow{2}{*}{Polishing} & Special mirror wax & 2.8$\pm$0.2 & 2.9$\pm$1.3 & 5.8$\pm$0.3 &3.8$\pm$0.9 \\ \cline{2-6}
    & One step fast wax & 3.1$\pm$0.3 & 5.0$\pm$1.7 & 5.5$\pm$0.4 & 34.5$\pm$3.1 \\  \hline
    \multirow{2}{*}{Alconox powder} & - & $<$0.37 & $<$4.4 & $<$0.24 & 5.9$\pm$1.3 \\ 
    \cline{2-6}
    & Measured at CJPL & 0.23$\pm$0.02 & 0.82$\pm$0.14 & 0.05$\pm$0.02 & 7.5$\pm$1.0 \\ \hline
    \multirow{2}{*}{Sand paper} & Black base-plate & 7.2$\pm$0.6 & $<$21 & 13.6$\pm$0.6 & $<$3.7 \\ \cline{2-6}
    & Green base-plate & $<$1.7 & $<$10.2 & $<$1.6 & $<$2.2 \\  \hline
    \multirow{2}{*}{Protection film} & Glue on PE film & $<$0.2 & $<$2.3 & $<$0.1 & $<$0.6 \\ 
    \cline{2-6}
    & Water soluble film  & $<$0.47 & $<$10.8 & 0.58$\pm$0.19 & $<$1.56 \\ 
    \hline 
    \end{tabular}
\end{table}

\section{Method for measuring U/Th in acrylic by ICP-MS}
\label{sec3}
Different from gamma spectroscopy, the principle of the Inductively Coupled Plasma Mass Spectroscopy (ICP-MS) is to detect the nuclei according to their exact atomic weight. ICP-MS can measure the isotopes $^{238}$U and $^{232}$Th directly without other information on the chain, and the sensitivity can reach the sub-ppt level. However, the sample measured by ICP-MS must be in liquid form, so a careful and optimized pre-treatment on the sample is needed. 

\begin{figure}[htbp]
\centering
    \includegraphics[width=8cm]{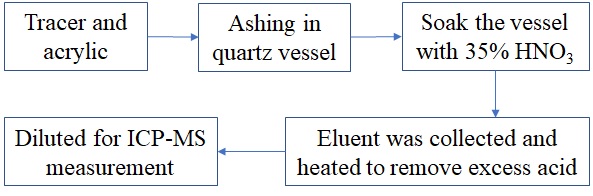}
    \caption{The flow chart for the pre-treamtent of acrylic samples. The ashing of the acrylic sample is done with the microwave muffle furnace described in the text.}
    \label{fig:flowchart}
\end{figure}

For the radio purity measurement on acrylic surface samples, we have developed a new method based on microwave ashing and ICP-MS equipment. The design for the pre-treatment flow is shown in Fig.~\ref{fig:flowchart}. Both the acrylic sample and the tracers (the natural-non-existing nuclei $^{229}$Th and $^{233}$U) are ashed by the machine in a quartz vessel. The residual is collected by soaking with 35\% HNO$_{3}$. The eluent is collected and heated to remove the excess acid, and the solution is finally diluted for the ICP-MS measurement.

In Ref.~\cite{Cao:2020zyr}, a different approach was used: the acrylic was vaporized and burned in a two-stage furnace, while clean gas (mixed N$_2$/O$_2$) was supplied to the furnace. Monitoring of the pressure in the whole system is essential in this case. Compared with the pre-treatment in~\cite{Cao:2020zyr}, with the new technique there is no need for gas input to the microwave muffle furnace, and the overall operation is easier and safer. All the pre-treatment process is done in a class 10,000 tent, while the ICP-MS measurement is performed in a class 1,000 room. The cleanliness of the quartz vessels is quite important for the sensitivity of the measurement, and their cleaning procedure  is similar to that described in Ref.~\cite{Cao:2020zyr}. All the containers are soaked in the two-stage acid cylinders filled with 6~mol/L HNO$_3$ for at least one day at each stage to remove the U/Th on the surface. In the end, the containers are filled with 6~mol/L HNO$_3$ and boiled for 10~min, then they are rinsed with fresh water and ready for use.

\begin{figure}[htbp]
    \centering
    \includegraphics[height=4.5cm]{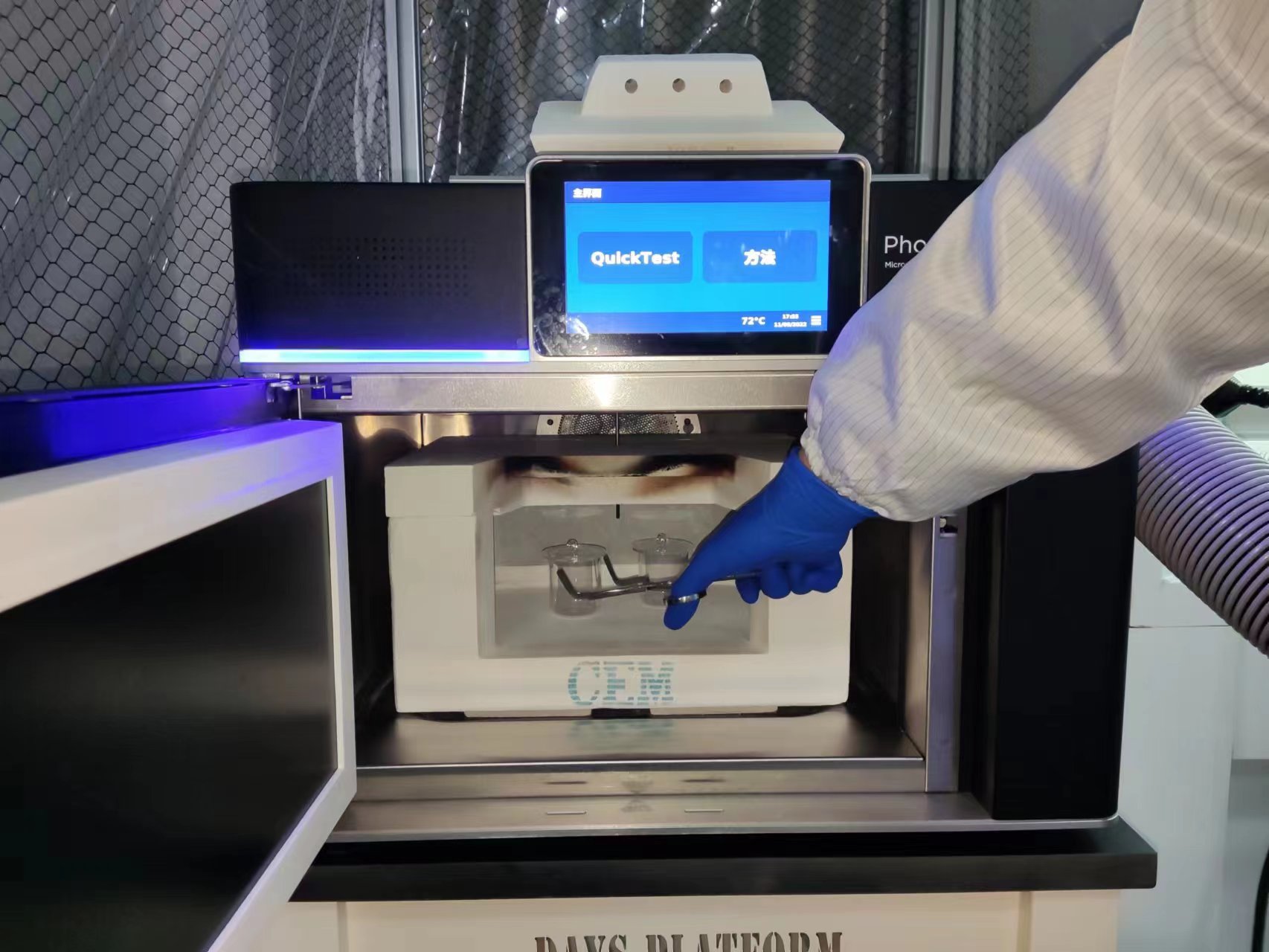}
    \qquad
    \includegraphics[height=4.7cm]{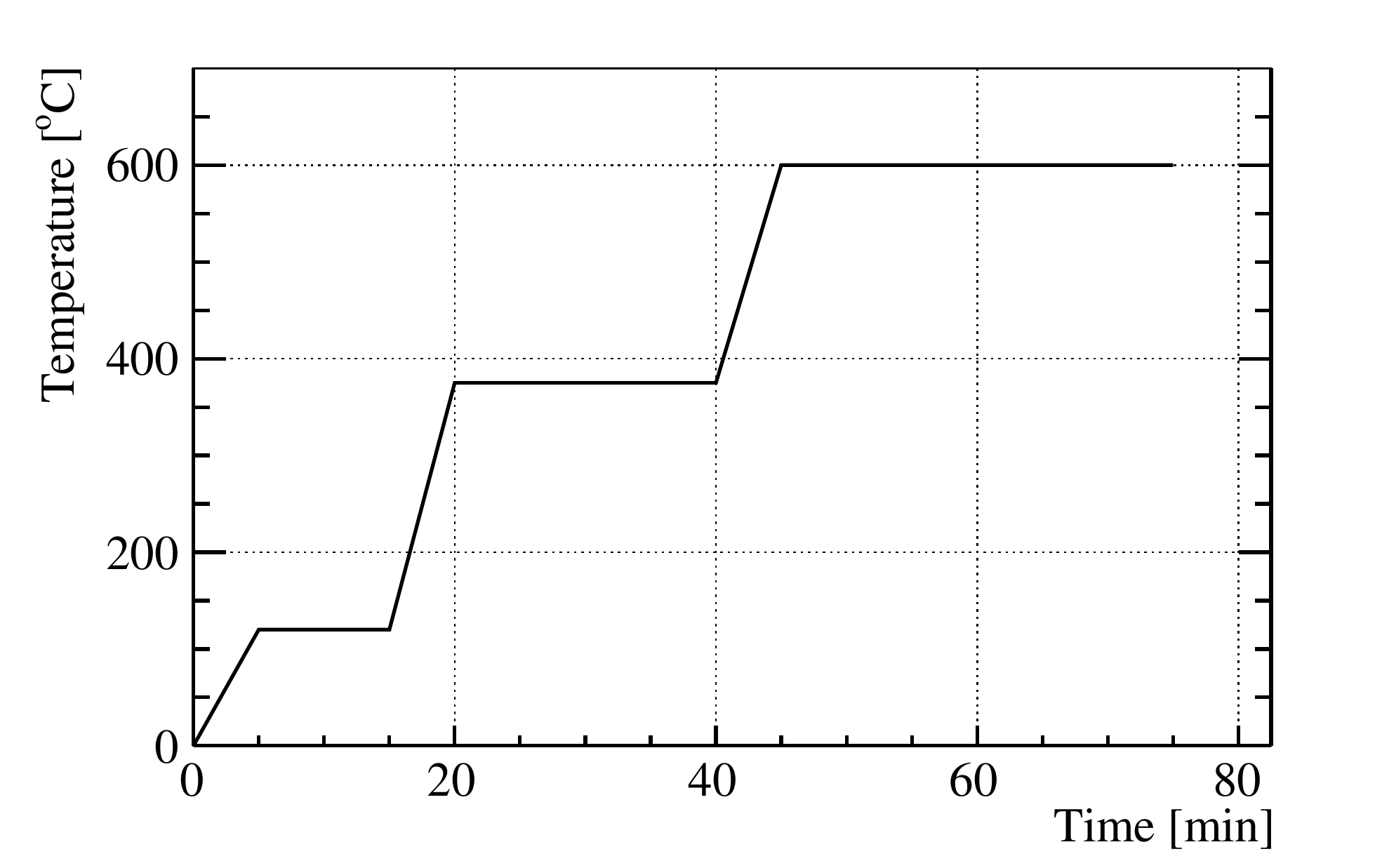}
    \caption{The microwave muffle furnace (Phoenix BLACK) used for the pre-treatment of acrylic is shown in the left figure. The optimized temperature as a function of time for acrylic ashing is shown in the right figure.}
    \label{fig:muffle}
\end{figure}

The microwave muffle furnace (Phoenix BLACK~\cite{muffle}) is put in the clean tent and used for ashing the acrylic sample, as shown on the left of Fig.~\ref{fig:muffle}. The temperature for ashing is optimized to achieve no visible organic residual in the vessel, and the optimized temperature as a function of time is shown on the right of Fig.~\ref{fig:muffle}. The quartz vessel is used as the sample container and well cleaned. The $^{229}$Th and $^{233}$U standards (preserved in 2\% acid) are added together with the acrylic sample to the vessel in the beginning, and the recovery efficiency can be evaluated by measuring the $^{229}$Th/$^{233}$U in the solution sent to ICP-MS. The average recovery efficiency of U/Th is calculated as (95$\pm$13)\% based on more than one hundreds measurements (the given uncertainty is the standard deviation). The recovery efficiency is calibrated every time a sample is measured, and the single precision is better than 5\%. To estimate the background for the pre-treatment, we have followed the same procedure without the acrylic sample (blank test). With careful background control, the blank test showed that the absolute background for the pre-treatment amounts to 0.24$\pm$0.07~pg for $^{238}$U and 0.38$\pm$0.04~pg for $^{232}$Th.

\section{Study of the procedure for acrylic surface treatments}
\label{sec4}
\subsection{Solution absorption in acrylic}
\label{sec4.1}
\begin{figure}[htbp]
    \centering
    \includegraphics[width=8cm]{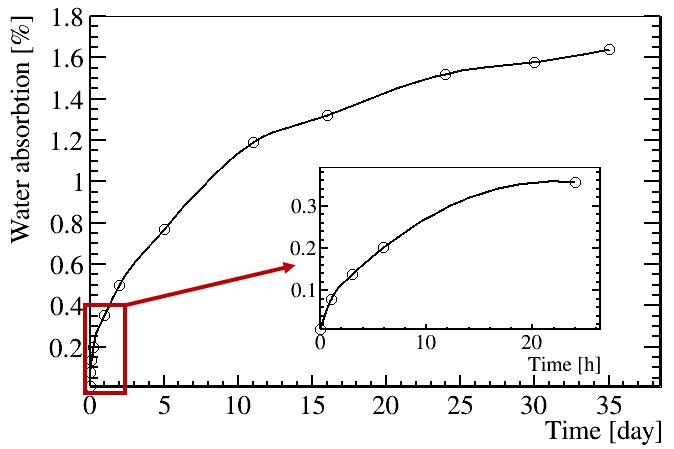}
    \caption{ Results of water absorption tests performed on flat acrylic samples with 2 mm thickness. The mass proportion of absorbed water to acrylic as a function of time is shown during 35 days, and the data points in the first day are magnified in the sub-figure.}
    \label{fig:waterAbsorb}
\end{figure}

The acrylic panel is cleaned after surface treatments at the company before shipment, and the whole acrylic sphere is finally cleaned onsite after installation. Acrylic can absorb water, thus traces of radioactivity in cleaning water may diffuse into acrylic. The absorption of water in acrylic obeys the mathematical laws of diffusion. We have performed several water absorption tests on flat acrylic samples with 2~mm thickness. The mass proportion of absorbed water in acrylic as a function of time is shown in Fig.~\ref{fig:waterAbsorb}. The absorption of water in acrylic did not reach equilibrium in one month, and the mass of absorbed water in acrylic increased almost linearly at the beginning, which is far away from the equilibrium state. Therefore, there is no impact of the acrylic panel thickness when exposure to water is shortened to several hours. Assuming we will clean the acrylic surface during one hour at the company, the absorbed water in acrylic can reach about 0.08\% for the sample with 2~mm thickness, which is 1~g/m$^2$ water absorption for one side of acrylic surface. 

For the cleaning procedure, we use Alconox solution (0.1\% of water) for degreasing and deionized water for rinsing. Traces of radioactivity in both water and Alconox can diffuse into acrylic during cleaning. We use deionized water with 10$^{-14}$~g/g U/Th to perform water cleaning, so the residual traces of radioactivity from absorbed water is negligible. Even though the Alconox is not as clean as the deionized water, most of Alconox can be removed by water rinsing. To validate how much radioactivity from Alconox can go into acrylic, we soaked the acrylic sample with 2~mm thickness in Alconox solution with different concentrations and time exposure, and the radiopurity of the sample for U/Th is measured by ICP-MS in the end. We have soaked the samples in Alconox solution for two months in order to enlarge the possible radioactivity contamination in the bulk of acrylic (2~mm thickness). The results of the tests for U/Th surface contamination are shown in Table~\ref{tab:soakTest}.

\begin{table}[htbp]
\centering
\caption{ \label{tab:soakTest} The acrylic samples with 2~mm thickness are soaked in different solutions with different time exposure. The U/Th contamination from solution absorption in acrylic after absorption is measured by ICP-MS. }
\smallskip
\begin{minipage}[c]{\textwidth}
  \resizebox{\textwidth}{!}{
	\begin{tabular}{c|c|c|c|c|c}
        \hline
    No. & Sample preparation & Exposure & Mass[g] & $^{238}$U[ppt] & $^{232}$Th[ppt] \\ \hline
    1 & No soaking & - & 3.25 & 0.5$\pm$0.1 & 1.7$\pm$0.1 \\ \hline
    2 & 0.1\% Alconox solution & 1 day & 3.21 & 0.6$\pm$0.1 & 2.3$\pm$0.2 \\ \hline
    3 & 0.1\% Alconox solution & 2 months & 3.18 & 1.9$\pm$0.1 & 2.0$\pm$0.3 \\ \hline
    4 & 2\% Alconox solution & 2 months & 3.74 & 4.4$\pm$0.2 & 2.3$\pm$0.2 \\ \hline
    5 & 0.1\% Alconox solution + 30\% HNO$_3$ & 2 months + 1 day & 3.11 & 0.5$\pm$0.1 & 1.5$\pm$0.2 \\ \hline
    6 & Tap water after filter [739 ppt $^{238}$U, 0.02 ppt $^{232}$Th] & 1 day & 3.28 & 3.0$\pm$0.1 & 1.7$\pm$0.2 \\ \hline
    7 & 1\% HNO$_3$ [106.7 ppt $^{238}$U, 614.5 ppt $^{232}$Th] & 1 day &  3.18 & 2.9$\pm$0.2 & 32.2$\pm$1.5 \\ \hline
    \end{tabular}}
    \end{minipage}
     \end{table}

\begin{itemize}
\item No.1-5: we can see a clear increase on $^{238}$U in acrylic for Alconox solutions with different concentration and exposure. However, the increase of $^{232}$Th is not obvious due to the lower $^{232}$Th radioactivity in Alconox powder as shown in Table~\ref{tab:HPGe}. Even though the Alconox can be removed by further deionized water rinsing, the radioactivity from Alconox can diffuse into acrylic together with water. If the residual radioactivity stick on acrylic from Alconox solution is due to the active agent, the residual can not be easily removed by cleaning or acid. On the contrary, if the mechanism of residual consists in ionic diffusion from Alconox solution to acrylic, such residual can be removed by acid. We observed that the radioactivity absorbed in acrylic can be further removed by acid soaking, as shown in result No.5. 

\item No.6-7: To further validate that the diffusion of radioactivity from Alconox to acrylic is in ionic form, we have soaked another two acrylic samples in two kinds of water. One is tap water after 0.2~$\upmu$m filter, and radioactivity exists in small particles. The other one is U/Th standard solution with 1\% HNO$_3$, and radioactivity is in ionic form. As shown in No.6 and No.7, the $^{238}$U concentration in acrylic reached equilibrium at 3~ppt, while $^{232}$Th can reach much higher to about 30~ppt. The reason for the difference between $^{238}$U and $^{232}$Th is their different intrinsic physicochemical properties, which is also discussed in~\cite{LaFerriere:2015owa}. Thorium is more reactive than uranium, thus it is easier for thorium to stick to acrylic. A similar phenomenon is observed in preparing tap water, most of the thorium is filtered and less residual is found in the water. From the results in No.6 and No.7, radioactivity can also diffuse into acrylic without the active agent, so the radioactivity in Alconox solution can diffuse into acrylic due to the diffusion of ion, not the active agent. The final residual on acrylic from the cleaning solution relies on the measurement of the surface sample.
\end{itemize}

\subsection{Contamination from different surface treatments}

To study the surface contamination, we have treated the surface with different procedure and directly measured the residual of radioactivity on surface. The surface samples were taken by scraping the surface with well cleaned erasing knife, and the thickness of the surface sample can reach 5-10~$\upmu$m. The radioactivity of the raw surface for one JUNO acrylic panel without any surface treatment is measured to be (52$\pm$1)~ppt $^{238}$U and (133$\pm$5)~ppt $^{232}$Th, which is 1-2 orders of magnitude higher than the bulk measurement. The radioactivity distribution along the depth was measured in Ref.~\cite{LA-ICPMS}, and has proven that U/Th contaminations can extend down to tens of $\upmu$m from the surface. In addition, the radon daughters ($^{210}$Pb, $^{210}$Po) deposited on the acrylic surface is usually non-negligible.  So we decide to remove at least 0.1~mm depth of acrylic and to maintain a high light transparency ($>$96\% at 420~nm ~\cite{Li:2021hzh} in ultrapure water environment) at the same time. 

A dedicated experiment was performed on a flat acrylic panel (not a JUNO panel) with a 1~m$^2$ area by applying different surface treatments including sanding, polishing, and cleaning successively. All of these treatments are done in a 10,000 class tent. The sanding of acrylic surfaces starts from 400 mesh (to remove at least 0.1~mm depth of acrylic) and is pursued by using sand papers with larger mesh, 800, 1200, 2000, and 3000. The higher the mesh number, the smoother the acrylic surface, and higher light transparency can be achieved. However, more steps increase the risk of contaminations. We have measured the surface sample with different mesh numbers of 1200, 2000, and 3000, and the results for residual U/Th concentration are consistent within 20\% among these samples. 

Polishing of acrylic is performed by a wool wheel with polishing fluid, and the typical fluid is purchased from Saint-Gobain~\cite{Norton}, whose components are organic. If there is polishing with mirror wax after sanding, sanding to 1200~mesh number is enough to reach the required light transparency. Part of the polishing fluid can be further removed by degreasing with Alconox solution. To quantify the contamination from residual polishing fluid and Alconox solution, we have done many tests on the acrylic panel with sanding to 1200~mesh number, and the results are shown in Table~\ref{tab:cleanTest}. The water cleaning is realized by a water jet with high pressure for 20 times, and the Alconox cleaning is done by spraying the Alconox on surface and wiping with a clean cloth for several times. Compared to the sample No.1, we have done additional cleaning with Alconox solution on No.2, and there is obvious radioactive residual on the surface from Alconox. For the samples No. 3 and No. 4, we have done polishing with the fluid for both samples, and performed cleaning with Alconox only on No.4. From the results, it is quite clear that a large amount of polishing fluid residual exists on the acrylic surface after cleaning even with Alconox solution. 

\begin{table}[htbp]\small
\centering
     \caption{ \label{tab:cleanTest} Many cleaning tests have been done on the non-JUNO acrylic panel after sanding to the 1200 mesh number. The water cleaning is realized by a water jet with high pressure 20 times, and the Alconox cleaning is done by wiping the surface with a clean cloth several times. The polishing is done with mirror wax. After the treatments, surface samples are taken by scraping the surface with well cleaned erasing knife, and the thickness of the surface sample can reach 5-10~$\upmu$m. The radioactivity in the surface samples is summarized in this table. }
\smallskip
	\begin{tabular}{c|c|c|c|c}
        \hline
    No. & Sample preparation & Mass[g] & $^{238}$U[ppt] & $^{232}$Th[ppt] \\ \hline
    1 & Water cleaning & 0.25 & 9.5$\pm$0.3 & 9.3$\pm$0.5 \\ \hline
    2 & Alconox + water cleaning & 0.34 & 27$\pm$1 & 22$\pm$2 \\ \hline
    3 & Polishing + water cleaning & 0.28 & 92$\pm$5 & 893$\pm$27 \\ \hline
    4 & Polishing + Alconox + water cleaning & 0.35 & 123$\pm$3 & 817$\pm$33 \\ \hline
    \end{tabular}
\end{table}

Based on the above measurements, polishing with mirror wax and Alconox cleaning have non-negligible residual on the acrylic surface, and we should avoid using them for surface treatments. To reach light transparency greater than 96\% at 420~nm in ultrapure water environment, we have performed sanding with 3000 mesh number, polishing with deionized water and, in the end, cleaning of the surface with a deionized water high pressure jet. We finally measured the surface of one JUNO acrylic panel following this procedure of surface treatments, with good results of (15.2$\pm$0.7)~ppt $^{238}$U and (24.3$\pm$0.7)~ppt $^{232}$Th in about 10~$\upmu$m depth, which is several times lower than the raw surface as shown in the beginning of this section. In addition, all the treatments of the inner surface of the acrylic panel were done in one day to avoid radon daughters deposition. 

\subsection{Protection film}

After surface treatments with sanding, polishing, and cleaning, the acrylic surface will be covered by a thin film to protect the acrylic from fallouts of radon and dust in the air during transportation and installation, which will last several months. There are two kinds of protection films that can be used to protect the acrylic surface, as discussed in Sec~\ref{sec2}.

Since the glue on the adhesive masking paper is soluble in water, we soaked the paper film in water for different exposure. By this way, we can measure the radiopurity of glue by ICP-MS with high precision. To reduce the effect of surface cleanliness in this measurement, we swab the surface of paper side with a little wet clean cloth. The paper is clipped out after soaking. Similar to pre-treatments in Figure~\ref{fig:flowchart}, the solution is vaporized and the residual is digested by acid. The results of the measurements are shown in Table~\ref{tab:film}. It seems that most of the glue is dissolved in water within one hour. The film put on inner surface of acrylic will be removed by high pressure water jet during the final cleaning. The paper film with water soluble glue on it is quite easy to be removed with water, so the contact time is much smaller than one hour. 

The mass of glue on the paper is 5~g/m$^2$, and the total mass of glue for the whole acrylic surface is 20~kg. Assume all the U/Th in glue stay on acrylic surface and further go into LS, the contamination to 20~kt LS is 10$^{-16}$~g/g, one order higher than our requirement of U/Th in LS (10$^{-17}$~g/g). In reality, almost all the glue can be dissolved in water when contacting with enough amount of water, and the radioactivity from glue can be absorbed in acrylic together with water. However, the flow rate of water is about 15~liters per minute, so the glue is highly diluted in water. Based on the study in Sec~\ref{sec4.1}, the water absorbed in acrylic can reach 1~g/m$^2$ after one hour soaking. In reality, the absorbed U/Th in acrylic from glue is quite small. 

\begin{table}[htbp]\small
\centering
 \caption{ \label{tab:film} The water soluble glue on paper film is digested in water solution and measured by ICP-MS. }
 \smallskip
	\begin{tabular}{c|c|c}
 \hline
    Exposure of soaking & 1 hour & 5 hours \\ \hline
    $^{238}$U [ppt] & 222$\pm$11 & 268$\pm$12 \\
    $^{232}$Th [ppt] & 117$\pm$5 & 81$\pm$5 \\
    \hline 
    \end{tabular}
\end{table}

Besides the direct radiopurity measurement of the glues, we have also performed the test on acrylic surface. We pasted the two kinds of films on acrylic surface, and took the surface sample after removing the film. We did not see an obvious difference between the samples with films and the raw panel sample, so it is safe to use both PE and paper film for acrylic protection. The whole acrylic sphere with 35.4~m in diameter is divided into 265 panels produced in the company separately. Considering the toughness of the films, we prefer to use PE film in the company, which is better for protection of the panels during transportation and installation. All of these panels will be bonded layer to layer from top to bottom at JUNO site. After bonding the panels of one layer, we will clean the surface of the corresponding layer and cover the inner surface with paper film. When the whole sphere will be finished, we will finally clean the inner surface with high pressure water jet, and the paper film will be easily removed without manual operation inside the sphere.

\section{Summary}
\label{sec5}
An acrylic vessel with 35.4 m in diameter is used as the 20~kt LS container for JUNO. The cleanliness of the inner acrylic surface is quite important to ensure an excellent LS radiopurity. To remove the contamination near the acrylic surface during production, many treatments will be done on the acrylic surface before shipment. From this study, we found the polishing mirror wax and Alconox solution have non-negligible U/Th residuals on the surface. The final recommended procedure for the surface treatments consists of sanding up to 3000 mesh number, polishing and cleaning with deionized water to achieve low background and high light transparency. Finally, a thin PE film will cover acrylic surface during transportation, and the film will be replaced by the adhesive masking paper for the inner surface after onsite bonding. This operation is quite important before finally removing the film by water jet, since the glue on the adhesive masking paper is water soluble. This procedure for acrylic surface treatments is also applicable to other low background experiments. In addition, a new method for pre-treatment of acrylic samples based on a microwave muffle furnace is shown in this paper, and the sensitivity of the measurement with ICP-MS can reach sub-ppt for U/Th in acrylic.

\acknowledgments
This work is supported by the Youth Innovation Promotion Association of the Chinese Academy of Sciences, the National Natural Science Foundation of China (No. 11905226), and the Strategic Priority Research Program of the Chinese Academy of Sciences (Grant No. XDA10010200).

\bibliographystyle{apsrev4-1}
\bibliography{reference}

\end{document}